\documentclass[twocolumn,preprintnumbers,amsmath,amssymb,pra,epsf]{revtex4}
\usepackage{graphicx}
\usepackage[dvips]{color}
\begin{document}

\title{Locomotion of microspheres for super-resolution imaging}
\author{Leonid A. Krivitsky$^{1}$, Jia Jun Wang$^1$,Zengbo Wang,$^2$ and Boris Lukiyanchuk$^1$}
\email{Leonid_Krivitskiy@dsi.a-star.edu.sg}
\affiliation{$^1$Data Storage Institute, Agency for Science Technology and Research, 5 Engineering Drive I, 117608 Singapore
\\$^2$School of Electronic Engineering, Bangor University, Dean Street, Bangor LL57 1UT, Gwynedd, UK} \vskip 24pt

\begin{abstract}
\begin{center}\parbox{14.5cm}
{Super-resolution virtual imaging by micron sized transparent beads (microspheres) was recently demonstrated by Wang \textit{et al.}. Practical applications in microscopy require control over the positioning of the microspheres. Here we present a method of positioning and controllable movement of a microsphere by using a fine glass micropipette. This allows sub-diffraction imaging at arbitrary points in three dimensions, as well as the ability to track moving objects. The results are relevant to a broad scope of applications, including sample inspection, microfabrication, and bio-imaging.}
\end{center}
\end{abstract}

\maketitle\narrowtext

\section{Introduction}

Development of imaging techniques with resolution beyond the diffraction limit (super-resolution) is of utmost importance for scientific community. Several super-resolution techniques have been implemented up to date, including: stimulated emission depletion microscopy (STED), structured illumination microscopy (SIM), ‎and photo activated localization microscopy (PALM) [1]. However, the price tags for commercially available systems range at around 1 million dollars, they require specific sample preparation, and involve sophisticated image processing algorithms. Scanning probe techniques, such a near-field scanning optical microscope (NSOM), may represent a more accessible alternative. However, they suffer from efficiency [2], and rather slow acquisition times. Therefore, development of low-cost and easily implementable solutions for super-resolution imaging is highly desirable.

One such promising solution is the microsphere nanoscopy technique, pioneered by us. The technique uses transparent beads with typical diameters of few microns (microspheres) are used as near-field to far-field lenses [3]. The microsphere focuses incident light to a subwavelength spot, and transfers a near-field image of the close-contacting object to a virtual image on the opposite side of the microsphere. The virtual image is unaffected by diffraction, and can be observed by adjusting a focal plane of a microscope objective lens. The experimentally demonstrated resolution of the technique is 50 nm. The technique has also been demonstrated in a liquid environment, which is highly appealing for biological applications [4]. 

Practical implementation of the technique requires high precision manipulation of the microsphere. First, the manipulation shall be able to position the microsphere at a specific location on the sample surface, which is essential for sample inspection applications. Second, it shall also be able to move the microsphere to track a moving object, which is essential for biological applications, such as observation of live bacteria and viruses. Optical tweezers have been used for locomotion of microspheres both for imaging and nano-focusing applications [5-8]. However, such setups may be costly, as they require additional components (light source, focusing objective lens, and dichroic optics). Optical tweezers cannot be used when samples are non-transparent at the wavelength of the trapping laser, and their operation in air is quite challenging. An alternative approach exploits random Brownian motion of imaging spheres in the vicinity of the object. A specific image processing algorithm is used to reveal a super-resolved image [9]. However it is essential that a  drop of liquid is deposited on the sample surface, which may not always be desirable for sample inspection applications.

In this work, we introduce an accessible and cost effective method of positioning and controlling movement of individual microspheres by attaching them to a sharpen tip of a rigid shaft (glass rod) using air suction, or UV-curable glue. The shaft is mounted on a three dimensional translation stage. By driving the translation stage, the microsphere, attached to the shaft, can be moved to any desired position in three dimensional space. The inclusion of the shaft does not effect the super-resolution imaging capabilities of microspheres, although they are stuck together. This work provides a new technology platform for a variety of super-resolution applications based on microspheres, including imaging, single-molecule detection, and laser nano-patterning.

\newpage

\section{Results}

We tested our method with two samples: one with sub-200nm feature and the other with sub-100 nm feature. The first sample is a polymer grating structure, fabricated by laser interference lithography on a silicon substrate. The scanning electron microscope (SEM) image of the structure is shown in Fig.1a, and an optical image is shown in Fig.1b. The width of the stripes is about 170 nm with 550 nm gaps, and the height of the grooves is about 500 nm.

The second sample is composed of gold squares, deposited next to each other on the silicon substrate. It is fabricated using electron beam lithography. The squares of 500x500 nm$^2$ are separated by a 73 nm gap, and have a height of 50 nm. The pairs of the split-squares have a 5 $\mu$m pitch in X- and Y- directions. The SEM image of one split square is shown in Fig.2a, and the microscope image is shown in Fig.2b.

\subsection{Demonstration of the positioning of the microsphere with the polymer sample}

Media 1 of Supplementary materials shows movement of the microsphere across and along the grating (X-, and Y-directions in Fig.4). The sphere is attached to the pipette tip using vacuum suction. The contact is proven to be tight, since the microsphere does not chirp off during the scan. Fig.1c shows a still virtual image of the structure, magnified by the probe in approximately 6 times. The magnified virtual image is converted to gray scale, and pixel values are retrieved using Mathematica. The corresponding intensity profile is shown in Fig.1d. The positioning accuracy of the probe is limited by the resolution of the manipulator (~20 nm). As it was mentioned above, it is also possible to fix the position of the probe and move the microscope stage instead. In this case the accuracy of the positioning of the probe is provided by the resolution of the microscope stage (~5 nm).

\begin{figure}
\centering
\includegraphics*[width=\linewidth]{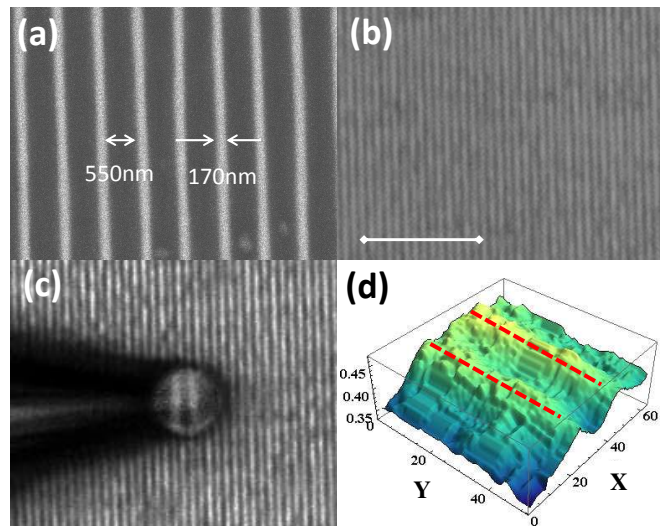}
\caption{(a) SEM image of the polymer sample deposited on a silicon substrate (b) Optical microscope image of the sample in air under 50x objective lens. The scale bar is 10 $\mu$m. (c) Virtual image of the structure in air magnified by the probe with a 50x objective lens. (d) Intensity profile of the magnified virtual image from a white dashed area in (c). X-, and Y-axes scales in (d) correspond to pixel numbers.}
\label{test}
\end{figure}

\subsection{Imaging of split-squares structure using the microsphere}

In order to demonstrate relevance of the method to sample inspection applications, we use the sample with gold split-squares with 73 nm gap, see Fig.2a. Optical image of the sample, see Fig.2b, does not allow resolving the gap between the squares. The accurate positioning of the microsphere between the squares using the micropipette is shown in Media 2 of Supplementary materials. Using the micromanipulator for movement of the microsphere allows to visualize the gap, as well as the borders of the structure. The observed virtual image of the gap is shown in Fig.2c. The corresponding intensity profile, obtained by the same algorithm as in Fig.1d, is shown in Fig.2d. In Fig. 2d the gap between the squares can be clearly observed.

\begin{figure}
\centering
\includegraphics*[width=\linewidth]{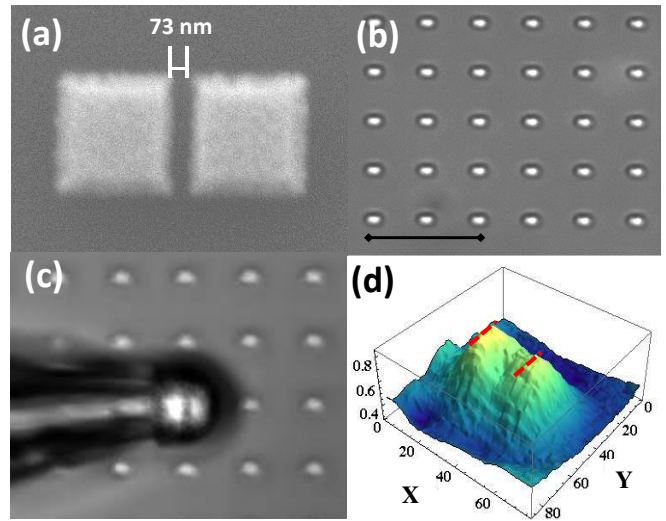}
\caption{(a) SEM image of the gold split-squares nanostructure. (b) Optical microscope image of the structure in air under 50x objective lens. The scale bar is 10 $\mu$m. (c) Magnified virtual image of the structure in air by the microsphere under 50x objective lens. (d) Intensity profile of the magnified virtual image from a white dashed area in (c). X-, and Y-axes scales correspond to pixel numbers.}
\label{test}
\end{figure}

\section{Discussion}

To understand the influence of micropipette on the microsphere imaging, numerical simulations were performed with CST Microwave Studio for following cases at 600 nm wavelengths: sole microsphere and microsphere with micropipette. The results shown in Fig.3 reveal that the micropipette has almost a negligible effect on the near-field focusing under the particle. In both cases, the focus spot size is about ~0.58$\lambda$, close to the diffraction limit; and the focal length is in a subwavelength scale: ~0.92$\lambda$ away from the particle bottom. Due to contacting nature of particle and substrate, strong near-field interaction between them will take place, leading to the conversion of surface evanescent waves into propagating waves reaching the far-field with ultrahigh resolution. Such case is similar to those in NSOM [2]. The difference is that the efficiency of conversion, in the case of NSOM is typically $10^{-5}-10^{-6}$ while in the case of the microsphere the efficiency is close to unity.

Due to the diffraction one can expect resolution of $\Delta\lambda= \lambda/2n$, which for n=1.5 and $\lambda$= 600 nm is equal to 200 nm. Our experiment, as well as the earlier studies [3] exceeds this resolution (3-4 times). It is related to specific of the virtual imaging, which permits to obtain far-field imaging at the distance $R >> L > \lambda$ , where L is the object size. For this case presentation of the evanescent wave in the form of plane wave is not valid, and the resolution limit can be found from the information theory [11]. At typical experimental conditions this limits is about $\lambda/10$-$\lambda/17$, which is consistent with our observations. 

It is possible to combine our approach with image processing techniques based on moving imaging spheres, by enabling mechanical vibrations to the shaft holding the microsphere [9]. Application of algorithmic imaging processing techniques such as Coherent Diffractive Imaging  should help to improve the resolution even further [12].

\begin{figure}
\centering
\includegraphics*[width=\linewidth]{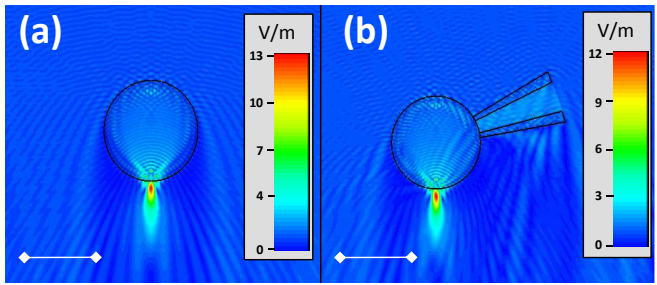}
\caption{Electric field distribution ($|E|$) around (a) the single microsphere and (b) the microsphere micropipette combo. The micropipette has negligible effect on the near-field focusing under the particle. The scale bar is 5 $\mu$m.}
\label{numerical}
\end{figure}

In conclusion, we improved the virtual image super-resolution technique by developing a method of controllable movement of imaging microspheres. The method allows controlling the position of the microsphere in 3 dimensions with high precision with no sacrifice of optical resolution. It represents a simple alternative to optical tweezers. It is important for applications, where accurate positioning of the optical probe is of crucial importance. These applications include, but not limited to quality control and inspection of samples, laser nano-patterning, tracing of moving biological objects such as living cells, viruses, and bacteria. 

\section{Methods}

\subsection{Optical setup}

We use an upright metallurgical microscope (BXFM, Olympus) mounted on top of an anti-vibration optical table, see Fig. 4. The microscope is equipped with a white light source (halogen lamp, LG-PS2, Olympus) and a color CMOS camera (SC30, Olympus). In the described experiments we use a 50X objective lens (0.55 NA, M Plan Apo, Mitutoyo) with 13 mm working distance. The sample is placed on a 3 dimension piezo translation stage with 5 nm resolution along X, Y, Z axis (NanoMax-TS, Melles Griot). We use dry silica microspheres (6.1 $\mu$m $\pm$ 0.59 $\mu$m diameter, SS06N, Bangs lab). To allow precise movement of a microsphere we attach it to a tip of a glass micropipette using air suction or a UV-curable optical glue (NOA81, Thorlabs). The pipette is mounted in a precision 3 dimensional manual flexure stage with 20 nm resolution in X, Y, Z axis (MDE 122 with MDE 216 HP adjusters, Elliot Scientific), referred to as a micromanipulator. The micropipette forms an angle of approximately 20 degrees to the sample surface. Note that there is no strict requirement to use the high resolution piezo stage for movement of the sample or the micropipette. The microsphere can be moved relative to the fixed sample with a simple mechanical manipulator. In such  simplified arrangement the estimated price of the components (excluding the price of the microscope) is approximately 2200 USD.

\begin{figure}
\centering
\includegraphics*[width=\linewidth]{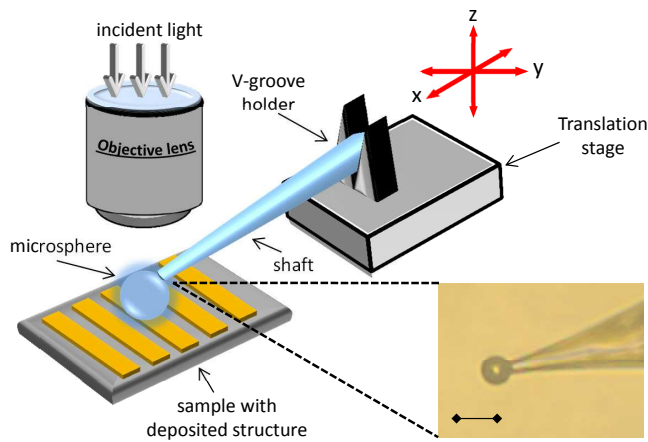}
\caption{Layout of the optical setup (not to scale). A microsphere lies on a sample and viewed under a microscope. It is attached to the micropipette using air suction or optical glue. The micropipette with the microsphere is attached to the micromanipulator via a V-grove holder. Driving the micromanipulator allows movement of the sphere in three dimensions. Inset shows the microscope image of the microsphere attached to the pipette by an optical glue. Image in air under 40x objective lens, the scale bar is 10 $\mu$m.}
\label{test}
\end{figure}

\subsection{Micropipette fabrication}

The micropipettes are pulled from borosilicate capillary glass (outer diameter 1.5 mm, inner diameter 1.12 mm, TW150-6, World Precision Instrument) by a micropipette puller (model P-97, Sutter Instruments). The parameters of the puller are set that the pulled micropipettes have opening diameters of 1 $\mu$m - 2 $\mu$m and taper lengths of about 3 mm [10]. The fabricated micropipette is fixed in a metallic hollow rod-shape adapter, which is attached to the micromanipulator via a V-groove holder(KM100V, Thorlabs). The plastic air tubing is channeled through the adapter and tightly fit to the opposite end of the micropipette. The end of the tubing is connected to a standard 10 ml syringe. 

\subsection{Imaging with microspheres}

In the experiment, a small quantity of microspheres is dispersed on the sample surface, and observed with the microscope. The tip of the micropipette is brought close (~1 $\mu$m - 2 $\mu$m) to a single microsphere using the micromanipulator. Negative air pressure in the micropipette is created by the pulling the plunger of the syringe, attracts the microsphere and attaches it to the tip. If the UV-curable optical glue is used, then, as soon as the microsphere touches the glue, a UV exposure is applied to strengthen the contact, see Fig.4. Then the pipette with the microsphere is lifted upwards and the microscope stage is moved to visualize the nanostructure. The microsphere is lowered, so that it “lands” on the nanostructure, see Fig.4. Through the microscope, one sees an image of the micropipette tip, the microsphere attached to it, and the nanostructure at the same time. The virtual image of the nanostructure is observed by adjusting the focal plane of the objective lens [3]. The sphere can be positioned at any desired location by using the micromanipulator. Note that the approach allows positioning the microsphere at the point above the sample (Z-direction in Fig.4). This provides an opportunity to study samples at various depths.

\section*{Acknowledgments}

We would like to thank Janaki DO Shanmugam for preparation of the samples, Reuben Bakker, Arseniy Kuznetsov, and Guillaume Vienne for valuable discussions. The work was supported by the Joint Council Office grant (Project No: 1231AEG025), and SERC Metamaterials Program on Superlens (grant no. 092 154 0099).


\begin{thebibliography}{00}

\bibitem{1} Schermelleh, L., Heintzmann, R., Leonhardt, H. A guide to super-resolution fluorescence microscopy, \textit{J. Cell Biol.} \textbf{190}, 165–175 (2010).
\bibitem{2}  Novotny, L.,  Hecht, B. \textit{Principles of Nano-Optics} (Cambridge University Press, 2006). 
\bibitem{3}  Wang, Z. \textit{et al.} Optical virtual imaging at 50 nm lateral resolution with a white-light nanoscope, \textit{Nature Communications} \textbf{2}, 1-5 (2011).
\bibitem{4}  Hao, X., Kuang, C., Liu, X., Zhang, H., Li, Y. Microsphere based microscope with optical super-resolution capability, \textit{Appl. Phys. Lett.} \textbf{99}, 203102 (2011).
\bibitem{5}	 Yakovlev, V. V., Luk’yanchuk, B. Multiplexed Nanoscopic Imaging, \textit{Laser Physics} \textbf{14}, 1065–1071 (2004).
\bibitem{6}  Faustov, A., Shcheslavskiy, V., Petrov, G. I., Lukyanchuk, B., Yakovlev, V. V. Highly multiplexed scanning nanoscopic imaging,\textit{ Proceedings SPIE “Nano/biophotonics”} \textbf{5331}, 21-28 (2004).
\bibitem{7}  Banas, A. \textit{et al.} Fabrication and optical trapping of handling structures for re-configurable microsphere magnifiers, \textit{Proc. SPIE} \textbf{8637}, Complex Light and Optical Forces VII, 86370Y (2013).
\bibitem{8}	 McLeod, E., Arnold, C. B. Subwavelength direct-write nanopatterning using optically trapped microspheres, \textit{Nature nanotechnology} \textbf{3}, 413-417 (2008).
\bibitem{9} Gur, A., Fixler, D., Mico, V., Garcia, J., Zalevsky, Z. Linear optics based nanoscopy, \textit{Opt. Exp.} \textbf{18}, 22222-22231 (2010).
\bibitem{10}	Sutter Instrument Company \textit{Pipette cook book}, www.sutter.com
\bibitem{11}	Narimanov, E. E. The resolution limit for far-field optical imaging, \textit{CLEO:2013 Technical Digest}, \textbf{QW3A.7} (2013)
\bibitem{12} Szameit A. \textit{et al.}, Sparsity-based single-shot subwavelength coherent diffractive imaging, \textit{Nature Materials} \textbf{11}, 455–459 (2012).
\end{thebibliography}
\end{document}